\documentclass[twocolumn,superscriptaddress,preprintnumbers,amsmath,amssymb,prl]{revtex4}
\UseRawInputEncoding
\usepackage{graphicx}

\begin{document}
\title{Casimir pressure in peptide films on metallic substrates: Change
of sign via graphene coating}

\author{
G.~L.~Klimchitskaya}
\affiliation{Central Astronomical Observatory at Pulkovo of the
Russian Academy of Sciences, Saint Petersburg,
196140, Russia}
\affiliation{Institute of Physics, Nanotechnology and
Telecommunications, Peter the Great Saint Petersburg
Polytechnic University, Saint Petersburg, 195251, Russia}

\author{
V.~M.~Mostepanenko}
\affiliation{Central Astronomical Observatory at Pulkovo of the
Russian Academy of Sciences, Saint Petersburg,
196140, Russia}
\affiliation{Institute of Physics, Nanotechnology and
Telecommunications, Peter the Great Saint Petersburg
Polytechnic University, Saint Petersburg, 195251, Russia}
\affiliation{Kazan Federal University, Kazan, 420008, Russia}

\author{
E.~N.~Velichko}
\affiliation{Institute of Physics, Nanotechnology and
Telecommunications, Peter the Great Saint Petersburg
Polytechnic University, Saint Petersburg, 195251, Russia}

\begin{abstract}
We find that the Casimir pressure in peptide films deposited on
metallic substrates is always repulsive which makes these films
less stable. It is shown that by adding a graphene sheet on top
of peptide film one can change the sign of the Casimir pressure
by making it attractive. For this purpose, the formalism of the
Lifshitz theory is extended to the case when the film and
substrate materials are described by the frequency-dependent
dielectric permittivities, whereas the response of graphene
to the electromagnetic field is governed by the polarization
tensor in (2+1)-dimensional space-time found in the framework
of the Dirac model. Both pristine and gapped and doped graphene
sheets are considered possessing some nonzero energy gap and
chemical potential. According to our results, in all cases the
presence of graphene sheet makes the Casimir pressure in peptide
film deposited on a metallic substrate attractive starting from
some minimum film thickness. The value of this minimum thickness
becomes smaller with increasing chemical potential and larger
with increasing energy gap and the fraction of water in peptide
film. The physical explanation for these results is provided,
and their possible applications in organic electronics are
discussed.
\end{abstract}

\maketitle

\newcommand{\kb}{{k_{\bot}}}
\newcommand{\ve}{{\varepsilon}}
\newcommand{\zy}{{(\zeta_l,y)}}
\newcommand{\yT}{{(y,T,\Delta,\mu)}}
\newcommand{\tP}{{\tilde{\Pi}}}
\newcommand{\tv}{{\tilde{v}_F}}
\newcommand{\tY}{{\tilde{Y}_l}}
\section{Introduction}

Considerable recent attention has been focused on organic materials
which combine electrical conductivity with high mechanical flexibility.
Carbon-based materials of this type have gained widespread acceptance
in organic electronics \cite{1} giving rise to creation of innovative
electronic devices, such as solar cells \cite{2,3}, field-effect
transistors \cite{4,4a,5,6}, light-emitting diodes \cite{7,8}, biomarkers
\cite{9,10,11}, etc. Many of the organic electronic devices employ thin
peptide films deposited on dielectric and metallic substrates for
their functionality \cite{12,13,14,15,16,17,18}. Physical phenomena
at organic-dielectric and organic-metal interfaces have been the
subject of much investigation (see, e.g.,
Refs.~\cite{19,20,21,22,23,24,25,26,27,28,29}.

One of these phenomena, which becomes essential in films of less than
1~$\mu$m thickness, is caused by the zero-point and thermal fluctuations
of the electromagnetic field. Fluctuations result in the van der Waals
and Casimir forces between two closely spaced materials layers separated
by a vacuum gap \cite{30,31} as well as in the internal free energies
and pressures induced in both freestanding and deposited on a substrate
films. The van der Waals and Casimir free energies and forces in layer
structures can be expressed via the frequency-dependent dielectric
permittivities of interacting layers by means of the Lifshitz theory
\cite{32}. In Refs.~\cite{33,34,35,36}, this theory was applied for
calculation of the van der Waals forces between films made of
organic materials. Based on the Lifshitz theory, the Casimir free
energy and pressure were also investigated for the freestanding and
deposited on substrates metallic and dielectric films
\cite{37,38,39,40,41,42}.

The fluctuation-induced free energy of both freestanding and deposited
on dielectric and metallic substrates peptide films was found in Ref.~\cite{43}.
For this purpose, a representation for the dielectric
permittivity of a typical model peptide along the imaginary frequency
axis was suggested using the results of Ref.~\cite{44} for electrically
neutral 18-residue zinc finger peptide in the microwave region and of
Ref.~\cite{45} for cyclic tripeptide RGD-4C in the ultraviolet region.
A behavior of the dielectric permittivity of typical peptide in the
region of infrared frequencies was modeled in the Ninham-Parsegian
representation.

It was shown that the fluctuation-induced free energy of peptide films
deposited on metallic substrates is positive, whereas for dielectric
substrates it can change its sign depending on the film thickness
\cite{43}. The effect of sign change arising in the case of dielectric
substrates was considered in more detail in Refs.~\cite{46,47}. According
to the obtained results, for sufficiently thick peptide films (thicker
than 135 nm for a film containing 10\% of water deposited on a SiO$_2$
substrate) the fluctuation-induced Casimir pressure becomes negative.
This corresponds to attraction of a film to a substrate and makes
film more stable. In Ref.~\cite{48} it was shown that the doping of
peptide film deposited on a dielectric substrate with metallic
nanoparticles results in a wider range of film thicknesses, where
the Casimir pressure is attractive, and further increases the film
stability.

In this paper, we consider peptide films deposited on metallic
substrates and show that the fluctuation-induced pressure is positive
which corresponds to a repulsive force and makes the peptide film
less stable. The doping of peptide with metallic nanoparticles is not
helpful in this case because it does not change the sign of the
Casimir pressure in films deposited on metallic substrates. In order
to make peptide films on metallic substrates more stable, we propose
to coat an additional graphene sheet on top of peptide film.

The fluctuation-induced Casimir pressure in a material layer
sandwiched between a sufficiently thick metallic plate and a
graphene sheet has not been investigated so far. This is a familiar
three-layer system where, however, one layer is two-dimensional.
Because of this, the Lifshitz formula for the Casimir pressure
should contain the standard reflection coefficients on the
boundary surface between peptide and metal (which are expressed via
the respective dielectric permittivities) and more sophisticated
ones on the boundary between peptide and graphene (which also depend
on the polarization tensor of graphene \cite{49}). Here, we present
the necessary formalism describing this system based on the first
principles of quantum electrodynamics at nonzero temperature.

Using the developed formalism, we have calculated the Casimir pressure
in pure peptide films and in peptide films containing different
fractions of water sandwiched between an Au substrate and a graphene
sheet with various values of the energy gap or chemical potential.
It is shown that the coating by a pristine graphene sheet on top of
pure peptide film makes the Casimir pressure in this film negative
for film thicknesses exceeding 211.7~nm. In so doing, for graphene
with a nonzero chemical potential the Casimir pressure becomes
attractive starting from smaller film thicknesses whereas for
graphene with a nonzero energy gap the same goal is reached for
thicker films.

Special attention is devoted to the most realistic case when a
peptide film contains some fraction of water and a graphene sheet
is characterized by the nonzero energy gap and chemical potential.
It is shown that in all cases the Casimir pressure in peptide film
becomes attractive for film thicknesses exceeding some definite
value wherein the pressure vanishes. This value depends on the
fraction of water in the film and on the values of the energy gap
and chemical potential of a graphene sheet. It is smaller for
peptide films with smaller fraction of water, larger chemical
potential and smaller energy gap of graphene. By contrast, for
larger fraction of water, smaller chemical potential and larger
energy gap of graphene the value of film thickness, such that the
Casimir pressure takes zero value, increases.

The obtained results could be useful for prospective devices of
organic electronics with further reduced dimensions to ensure
their stability.

The paper is organized as follows. In Sec.~II, the general
formalism is presented allowing calculation of the Casimir pressure
in thin film sandwiched between a metallic plate and a graphene
sheet. Section III is devoted to an impact of graphene coating
on the Casimir pressure in pure (dried) peptide film. The most
realistic case of peptide films containing some fraction of water
and graphene films possessing the nonzero energy gap and chemical
potential is considered in Sec.~IV. Section V presents our
conclusions and a discussion.

\section{Formalism for the Casimir pressure in a film sandwiched between metallic
plate and graphene sheet}

We consider the Casimir pressure in a peptide film of thickness $a$ characterized by the
frequency-dependent dielectric permittivity $\ve^{(1)}(\omega)$ deposited on metallic
substrate with the dielectric permittivity $\ve^{(2)}(\omega)$.
The sheet of graphene possessing the energy gap $\Delta$ and chemical potential $\mu$
is coated on the top of peptide film. This system is kept at temperature $T$ in thermal
equilibrium with the environment.

For the purpose of numerical computations, it is convenient to use the dimensionless
variables
\begin{equation}
\zeta_l=\frac{2a\xi_l}{c}, \quad
y=2a \left(k_{\bot}^2+\frac{\xi_l^2}{c^2}
\right)^{1/2}\!,
\label{eq1}
\end{equation}
\noindent
where $\xi_l=2\pi k_BTl/\hbar$ are the Matsubara frequencies, $k_B$ is the Boltzmann
constant, and $\kb$ is the projection of the wave vector on the plane of a film
(which is perpendicular to the Casimir force).

Then, the Casimir pressure in a peptide film is given by the Lifshitz formula \cite{30,31,32}
\begin{eqnarray}
&&
P(a,T)=-\frac{k_BT}{8\pi a^3}\sum_{l=0}^{\infty}{\vphantom{\sum}}^{\prime}
\int_{\zeta_l}^{\infty}\!yk^{(1)}\zy dy
\label{eq2}\\
&&~~~~~~
\times\sum_{\kappa}
\frac{1}{\left[R_{\kappa}\zy r_{\kappa}^{(1.2)}\zy\right]^{-1}
e^{k^{(1)}\zy}-1},
\nonumber
\end{eqnarray}
where
\begin{equation}
k^{(n)}\zy=\left[y^2+(\ve_l^{(n)}-1)\zeta_l^2\right]^{1/2}\!,
\label{eq3}
\end{equation}
\noindent
$\ve_l^{(n)}=\ve^{(n)}(i\xi_l)=\ve^{(n)}[ic\zeta_l/(2a)]$, $n=1,\,2$ for a peptide
and a substrate metal, respectively, a prime on the summation sign divides the term with
$l=0$ by 2, and the sum in $\kappa$ is over two polarizations of the electromagnetic
field, --- transverse magnetic ($\kappa={\rm TM}$) and
transverse electric ($\kappa={\rm TE}$).

The quantities $r_{\kappa}^{(1,2)}$ and $R_{\kappa}$ in Eq.~(\ref{eq2}) are
the reflection coefficients on the boundary planes between the peptide film and metallic
substrate and between the peptide film and graphene sheet, respectively.
Note that the metallic substrate is considered as a semispace (for this purpose its
thickness should be larger than 100~nm \cite{31}).

An explicit form of the coefficients $r_{\kappa}^{(1,2)}$ is well known.
These are the familiar Fresnel reflection coefficients calculated at the pure
imaginary Matsubara frequencies
\begin{eqnarray}
&&
r_{\rm TM}^{(1,2)}\zy=
\frac{\ve_l^{(2)}k^{(1)}\zy-\ve_l^{(1)}k^{(2)}\zy}{\ve_l^{(2)}k^{(1)}\zy+\ve_l^{(1)}k^{(2)}\zy},
\nonumber \\
&&
r_{\rm TE}^{(1,2)}\zy=
\frac{k^{(1)}\zy-k^{(2)}\zy}{k^{(1)}\zy+k^{(2)}\zy}.
\label{eq4}
\end{eqnarray}

The reflection coefficients $R_{\kappa}$ on the boundary plane between peptide and graphene are
more involved. If a peptide film and a graphene sheet are separated by a vacuum gap of width $d$,
the reflection coefficient takes the form \cite{31}
\begin{equation}
R_{\kappa}(\zeta_l,y;d)=\frac{r_{\kappa}^{(1,0)}\zy+r_{\kappa}^{({\rm gr})}\zy
e^{-dy/a}}{1+r_{\kappa}^{(1,0)}\zy r_{\kappa}^{({\rm gr})}\zy
e^{-dy/a}}.
\label{eq5}
\end{equation}
\noindent
Here, $r_{\kappa}^{(1,0)}$ are the Fresnel reflection coefficients on the boundary between
a peptide semispace and vacuum which are expressed via the dielectric permittivity of
peptide $\ve_l^{(1)}$ as
\begin{eqnarray}
&&
r_{\rm TM}^{(1,0)}\zy=
\frac{k^{(1)}\zy-\ve_l^{(1)}y}{k^{(1)}\zy+\ve_l^{(1)}y},
\nonumber \\
&&
r_{\rm TE}^{(1,0)}\zy=
\frac{k^{(1)}\zy-y}{k^{(1)}\zy+y}.
\label{eq6}
\end{eqnarray}

The reflection coefficients $r_{\kappa}^{({\rm gr})}$ in Eq.~(\ref{eq5}) are on a
freestanding in vacuum graphene sheet. These are the non-Fresnel coefficients whose
exact form is expressed in the framework of the Dirac model \cite{50} via the
polarization tensor of graphene in (2+1)-dimensional space-time
\begin{equation}
\Pi_{mn}(i\xi_l,\kb,T,\Delta,\mu)\equiv \Pi_{mn,l}(\kb,T,\Delta,\mu)
\label{eq7}
\end{equation}
\noindent
or, in the dimensionless form,
\begin{equation}
\tP_{mn,l}\yT =\frac{2a}{\hbar} \Pi_{mn,l}(\kb,T,\Delta,\mu),
\label{eq8}
\end{equation}
\noindent
where $m,\,n=0,\,1,\,2$.

The explicit expressions for the coefficients $r_{\kappa}^{({\rm gr})}$ are the following
(see Refs.~\cite{51,52} and also Refs.~\cite{53,54} where the dimensionless quantities
are used):
\begin{eqnarray}
&&
r_{\rm TM}^{({\rm gr})}\zy=\frac{y\tP_{00,l}\yT}{y\tP_{00,l}\yT+2(y^2-\zeta_l^2)},
\nonumber \\
&&
r_{\rm TE}^{({\rm gr})}\zy=-\frac{\tP_{l}\yT}{\tP_{l}\yT+2y(y^2-\zeta_l^2)},
\label{eq9}
\end{eqnarray}
\noindent
where the quantity
\begin{eqnarray}
&&
\tP_{l}\yT\equiv(y^2-\zeta_l^2){\rm tr}\tP_{mn,l}\yT
\nonumber \\
&&~~~~~~~~~~~~~~~~~
-y^2\tP_{00,l}\yT
\label{eq10}
\end{eqnarray}
\noindent
is expressed via the tensor trace ${\rm tr}\tP_{mn,l}\equiv\tP_{m,l}^{\,m}$.

Before presenting the exact expressions for the components of the polarization
tensor, we conclude that the reflection coefficients $R_{\kappa}$ entering Eq.~(\ref{eq2})
are obtained from Eq.~(\ref{eq5}) in the limiting case $d\to 0$:
\begin{eqnarray}
&&
R_{\kappa}\zy=\lim_{d\to 0}R_{\kappa}(\zeta_l,y;d)
\nonumber \\
&&
\phantom{R_{\kappa}\zy}
=\frac{r_{\kappa}^{(1,0)}\zy+r_{\kappa}^{({\rm gr})}\zy}{1+
r_{\kappa}^{(1,0)}\zy r_{\kappa}^{({\rm gr})}\zy},
\label{eq11}
\end{eqnarray}
\noindent
where $r_{\kappa}^{(1,0)}$ and $r_{\kappa}^{({\rm gr})}$ are defined in Eqs.~(\ref{eq6})
and (\ref{eq9}), respectively.

It only remains to present the expressions for $\tP_{00,l}$ and $\tP_l$.
We write out these expressions in the form allowing immediate analytic continuation
over the entire plane of complex frequencies \cite{55,56} and use the dimensionless
quantities \cite{53,54}.

We start with the case $l=0$. In this case the exact expressions for
$\tP_{00,0}$ and $\tP_0$ are given by \cite{53,54}
\begin{widetext}
\begin{eqnarray}
&&
\tP_{00,0}\yT=\frac{\alpha y}{\tv}\Psi(D_0)+\frac{16\alpha}{\tilde{v}_F^2}
\frac{ak_BT}{\hbar c}\ln\left[\left(e^{\frac{\mu}{k_BT}}+
e^{-\frac{\Delta}{2k_BT}}\right)\left(e^{-\frac{\mu}{k_BT}}+
e^{-\frac{\Delta}{2k_BT}}\right)\right]
\nonumber \\
&&~~~~~~~~~~~~~~~
-\frac{4\alpha y}{\tv}\int_{D_0}^{\sqrt{1+D_0^2}}du w_0(u,y,T,\mu)
\frac{1-u^2}{(1-u^2+D_0^2)^{1/2}},
\label{eq12} \\
&&
\tP_0\yT=\alpha\tv y^3\Psi(D_0)+4\alpha\tv y^3
\int_{D_0}^{\sqrt{1+D_0^2}}du w_0(u,y,T,\mu)
\frac{-u^2+D_0^2}{(1-u^2+D_0^2)^{1/2}}.
\nonumber
\end{eqnarray}
\end{widetext}
\noindent
Here, the following notations are introduced:
\begin{eqnarray}
&&
\Psi(x)=2[x+(1-x^2)\arctan x^{-1}],
\nonumber \\
&&
w_l(u,y,T,\mu)=\frac{1}{e^{B_lu+\frac{\mu}{k_BT}}+1}+
\frac{1}{e^{B_lu-\frac{\mu}{k_BT}}+1},
\nonumber \\
&&
B_l=B_l(y,T)=\frac{\hbar c}{4ak_BT}p_l(y),
\nonumber \\
&&
D_l=D_l(y)=\frac{2a\Delta}{\hbar c p_l(y)},
\nonumber \\
&&
p_l(y)=\left[\tilde{v}_F^2y^2+(1-\tilde{v}_F^2)\zeta_l^2
\right]^{1/2}\!,
\label{eq13}
\end{eqnarray}
\noindent
and $\tv=v_F/c\approx 1/300$, $\alpha=e^2/(\hbar c)\approx 1/137$ are the
dimensionless Fermi velocity for graphene and the fine structure constant,
respectively.

Now we consider the expressions for $\tP_{00,l}$ and $\tP_l$ with
$l\geqslant 1$. In this case it is unnecessary to use rather cumbersome exact
formulas because at $T=300~$K and film thicknesses $a>50~$nm it holds
$\zeta_l\gg\tv$ for all $l\geqslant 1$. As a result \cite{53,54,57}
\begin{eqnarray}
&&
\tP_{00,l}\yT\approx\frac{\alpha(y^2-\zeta_l^2)}{\zeta_l}\left[
\Psi(\Delta_l)+\tY\yT\right],
\nonumber \\[-1mm]
&&
\label{eq14} \\[-1mm]
&&
\tP_{l}\yT\approx{\alpha\zeta_l(y^2-\zeta_l^2)}\left[
\Psi(\Delta_l)+\tY\yT\right],
\nonumber
\end{eqnarray}
\noindent
where $\Delta_l=2a\Delta/(\hbar c\zeta_l)$ and the quantity $\tY$ is defined as
\begin{eqnarray}
&&
\tY\yT=2\int_{\Delta_l}^{\infty}\!\!du w_l(u,y,T,\mu)
\frac{u^2+\Delta_l^2}{u^2+1}
\label{eq15} \\
&&~
\approx2\int_{\Delta_l}^{\infty}\!\!du
\left(\frac{1}{e^{\pi lu+\frac{\mu}{k_BT}}+1}+
\frac{1}{e^{\pi lu-\frac{\mu}{k_BT}}+1}\right)
\frac{u^2+\Delta_l^2}{u^2+1}.
\nonumber
\end{eqnarray}
\noindent
In the last equation, we have taken into account that the major contribution
to Eq.~(\ref{eq2}) is given by $y\sim 1$. Computations show \cite{57} that
Eqs.~(\ref{eq14}) and (\ref{eq15}) lead to less than 0.02\% relative error
in the obtained Casimir pressures.

The reflection coefficients (9) on a graphene sheet with $l=0$ are simplified to
\begin{eqnarray}
&&
r_{\rm TM}^{({\rm gr})}(0,y)=\frac{\tP_{00,0}\yT}{\tP_{00,0}\yT+2y},
\nonumber \\
&&
r_{\rm TE}^{({\rm gr})}(0,y)=-\frac{\tP_{0}\yT}{\tP_{0}\yT+2y^3},
\label{eq16}
\end{eqnarray}
\noindent
where $\tP_{00,0}$  and $\tP_{0}$ are defined in Eq.~(\ref{eq12}).

By putting $l=0$ in Eq.~(\ref{eq6}), one obtains
\begin{equation}
r_{\rm TM}^{(1,0)}(0,y)=\frac{1-\ve_0^{(1)}}{1+\ve_0^{(1)}},
\quad
r_{\rm TE}^{(1,0)}(0,y)=0.
\label{eq17}
\end{equation}

Substituting Eqs.~(\ref{eq16}) and (\ref{eq17}) in Eq.~(\ref{eq11}), we arrive at
\begin{eqnarray}
&&
R_{\rm TM}(0,y)=
\frac{\tP_{00,0}\yT+y(1-\ve_0^{(1)})}{\tP_{00,0}\yT+y(1+\ve_0^{(1)})},
\nonumber \\
&&
R_{\rm TE}(0,y)=r_{\rm TE}^{({\rm gr})}(0,y).
\label{eq18}
\end{eqnarray}

For all $l\geqslant 1$ the reflection coefficients (\ref{eq11}) are obtained with
the help of Eqs.~(\ref{eq6}), (\ref{eq9}), and (\ref{eq14})
\begin{eqnarray}
&&
R_{\rm TM}\zy=\frac{k^{(1)}\zy\{\alpha[\Psi(\Delta_l)+\tY]y+\zeta_l
\}-\ve_l^{(1)}\zeta_ly}{k^{(1)}\zy\{\alpha[\Psi(\Delta_l)+
\tY]y+\zeta_l\}+\ve_l^{(1)}\zeta_ly},
\nonumber \\
&&
R_{\rm TE}\zy=\frac{k^{(1)}\zy-y-\alpha[\Psi(\Delta_l)+
\tY]\zeta_l}{k^{(1)}\zy+y+\alpha[\Psi(\Delta_l)+
\tY]\zeta_l},
\label{eq19}
\end{eqnarray}
\noindent
where $\Psi(\Delta_l)$ and $\tY$ are defined in Eqs.~(\ref{eq13}) and (\ref{eq15}).

Now that we have presented the required formalism, the Casimir pressure in a peptide film can be
computed by Eq.~(\ref{eq2}) using the reflection coefficients (\ref{eq4}), (\ref{eq18}) and
(\ref{eq19}).  Calculations of this kind are performed in the next sections for different
parameters of a peptide film and a graphene sheet.

\section{Impact of graphene coating on the Casimir pressure in dried peptide films}

In this section, we consider a somewhat idealized situation when the film is made of
pure dried peptide, which does not contain water, and graphene is pristine or is
characterized either by some nonzero chemical potential $\mu$ or by a nonzero energy
gap $\Delta$.

In the absence of graphene, the Casimir free energy of peptide films (both dried and
containing some fraction $\Phi$ of water) deposited on an Au substrate was found in
Ref.~\cite{43}. For this purpose, by combining available information about the optical
properties of two closely similar peptides in different frequency regions (see Sec.~I),
the values of dielectric permittivities of typical peptides $\ve_l^{(1)}$ containing
the fractions of water $\Phi=0$, 0.1, 0.25, and 0.4 were found at the pure imaginary
Matsubara frequencies $i\xi_0=0$, $i\xi_1,\,i\xi_2,\,\ldots,\,i\xi_{30}$
(see Fig.~2 of Ref.~\cite{43}). These values are used in all computations below.
It was shown \cite{43} that the Casimir free energy of peptide films with any $\Phi$
on an Au substrate is positive and decreases with increasing film thickness.

We begin with similar computations for the Casimir pressure performed by using
Eq.~(\ref{eq2}) where in the absence of graphene one should put
$r_{\kappa}^{({\rm gr})}\zy=0$ in Eq.~(\ref{eq11}) so that
\begin{equation}
R_{\kappa}\zy=r_{\kappa}^{(1,0)}\zy,
\label{eq20}
\end{equation}
\noindent
where $r_{\kappa}^{(1,0)}$ is defined in Eq.~(\ref{eq6}).

To compute the Casimir pressere in peptide film deposited on an Au substrate, one also
needs the values of dielectric permittivity of Au, $\ve_l^{(2)}$, at the imaginary
Matsubara frequencies. These are usually found by using the optical data for the complex
index of refraction of Au \cite{58} extrapolated down to zero frequency and the
Kramers-Kronig relations \cite{31}. It has been known that in calculations of the
Casimir force between metallic plates through a vacuum gap an extrapolation of the
optical data was made by means of either the dissipative Drude model or dissipationless
plasma model with diverged results. In so doing, contrary to expectations, theoretical
results using the plasma model extrapolation were confirmed experimentally and using
the Drude model extrapolation were experimentally excluded \cite{31,59,60}
(see also recent Ref.~(\cite{60a}). This is known as the Casimir puzzle \cite{61}.

In our case, however, both extrapolations lead to coinciding results because the TE
reflection coefficient on the boundary plane between peptide and vacuum vanishes at
zero frequency. In the presence of a graphene sheet on top of peptide film both
extrapolations lead to almost coinciding results because the TE
reflection coefficient on the boundary between graphene and vacuum at zero frequency
is very small.

\begin{figure}[!b]
\vspace*{-1.5cm}
\centerline{\hspace*{1.5cm}
\includegraphics[width=5.0in]{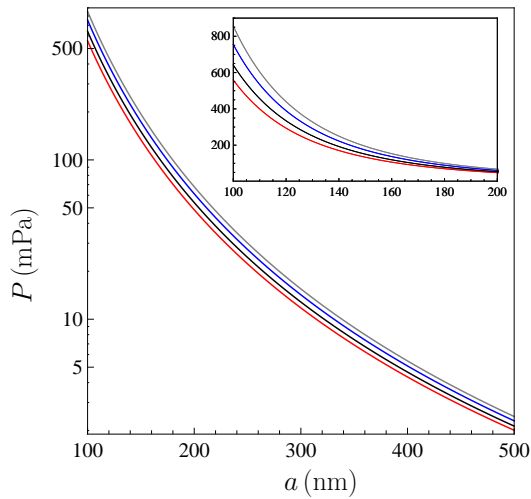}}
\vspace*{-10.5cm}
\caption{\label{fg1} The Casimir pressure in peptide films containing
$\Phi=0$, 0.1, 0.25, and 0.4 fractions of water deposited on an Au substrate
are shown as functions of film thickness in a logarithmic scale by the four
lines from bottom to top, respectively, at $T=300~$K. In the inset, the
region of smaller film thicknesses is shown in a uniform scale.
}
\end{figure}
In Fig.~\ref{fg1}, we present in the logarithmic scale computational results for
the Casimir pressure in peptide films deposited on Au substrate as the functions of
film thickness at room temperature $T=300~$K. The four lines from bottom to top are
for the dried films and for films containing $\Phi=0.1$, 0.25, and 0.4 volume fractions
of water. In the inset, the region of film thicknesses from 100 to 200~nm is shown
in a uniform scale. It is seen that in all cases the Casimir pressure remains positive,
i.e., the Casimir force is repulsive. Thus, the effects of electromagnetic fluctuations
make peptide coating less stable.

As is seen in Fig.~\ref{fg1}, the Casimir pressure decreases with increasing film
thickness and increases with increasing fraction of water in the film.
For instance, for the film of 150~nm thickness containing $\Phi=0$, 0.1, 0.25, and
0.4 fractions of water the Casimir pressure is equal to 133.8, 150.6, 173.6, and
194.1~mPa, respectively. We have checked that the doping of peptide film with
metallic nanoparticles which was used in order to increase film stability in the
case of dielectric substrate \cite{48}, is incapable to change the sign of the
Casimir pressure for metallic substrates.

Now we consider an effect of graphene coating on top of a dried peptide film.
We perform numerical computations of the Casimir pressure using Eqs.~(\ref{eq2}),
(\ref{eq4}), (\ref{eq18}), and (\ref{eq19}) for the cases of pristine graphene
which is gapless ($\Delta=0$) and undoped ($\mu=0$) and also for graphene
characterized either by some nonzero value of $\Delta$ (but undoped) or by
some nonzero value of $\mu$ (but gapless).

\begin{figure}[!b]
\vspace*{-3cm}
\centerline{\hspace*{1cm}
\includegraphics[width=6.50in]{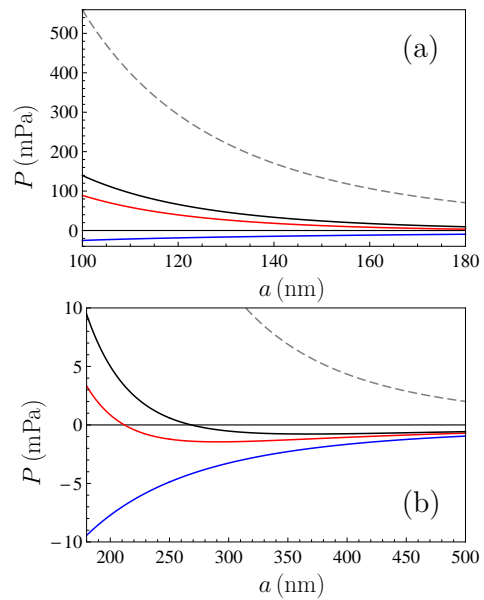}}
\vspace*{-13.1cm}
\caption{\label{fg2} The Casimir pressure in dried peptide film deposited on an Au
substrate and coated by a graphene sheet with ($\mu=0.25~$eV, $\Delta=0$),
($\mu=\Delta=0$), and ($\mu=0$, $\Delta=0.1~$eV) are shown as functions of film
thickness by the three solid lines from bottom to top, respectively, at $T=300~$K
in the regions of film thickness
(a) from 100 to 180~nm and (b) from 180 to 500~nm. The dashed line shows the Casimir
pressure in the absence of a graphene layer.
}
\end{figure}

In Fig.~\ref{fg2}(a,b) the computational results
for the Casimir pressure in graphene-coated dried peptide film
are shown as functions of the film thickness by the bottom, middle, and top solid lines
for graphene coatings with ($\mu=0.25~$eV, $\Delta=0$), ($\mu=\Delta=0$), and
($\mu=0$, $\Delta=0.1~$eV), respectively, at $T=300~$K in the regions of film thickness
(a) from 100 to 180~nm and (b) from 180 to 500~nm. The dashed line reproduces the
computational results obtained for a dried peptide film without graphene coating
(see the bottom line in Fig.~\ref{fg1}). The chosen values of $\mu$ and $\Delta$ are
typical for graphene sheets deposited on a substrate \cite{62}.

{}From Fig.~\ref{fg2}(a,b), it is seen that the presence of a graphene coating strongly
affects the Casimir pressure in peptide film. For films of any thickness, this pressure
becomes much smaller than in the absence of graphene coating. At some film thickness
$a_0$ the Casimir pressure vanishes and changes its sign from positive to negative
for thicker films. This happens at $a_0=268.5~$nm for the graphene coating with
$\mu=0$, $\Delta=0.1~$eV and at $a_0=211.7~$nm for a pristine graphene
[see the top and middle lines in Fig.~\ref{fg2}(b)].
For graphene with $\mu=0.25~$eV, $\Delta=0$
[the bottom lines in Fig.~\ref{fg2}(a,b)] we have $a_0<100~$nm.

In all cases the presence of graphene coating leads to attractive Casimir forces for
sufficiently thick peptide films and, thus, contribute to the film stability. In doing so
an increase of $\mu$ results in larger magnitudes of the negative Casimir pressures over
the wider region of film thicknesses, as compared to the case of pristine graphene,
whereas an increase of $\Delta$ leads to smaller in magnitude negative pressures than
for a pristine graphene and for thicker films. As an example, for a peptide film of
150~nm thickness the Casimir pressure is equal to 133.8~mPa in the absence of a graphene
sheet [see Fig.~\ref{fg1} and the dashed line in Fig.~\ref{fg2}(a)] but to 24.24, 12.35,
and --13.04~mPa in the presence of graphene sheets with ($\mu=0$, $\Delta=0.1~$eV),
($\mu=\Delta=0$), and ($\mu=0.25~$eV, $\Delta=0$),  respectively (see the top, middle and
bottom solid lines  in Fig.~\ref{fg2}(a)].

\begin{figure}[!h]
\vspace*{-3.3cm}
\centerline{\hspace*{1cm}
\includegraphics[width=5.0in]{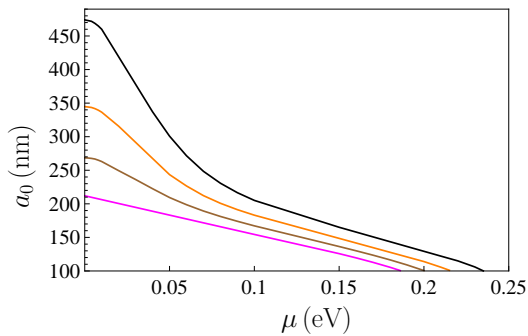}}
\vspace*{-10.6cm}
\caption{\label{fg3} The thicknesses of dried peptide films deposited on an Au
substrate and coated by a graphene sheet such that the Casimir pressure in the
film computed at $T=300~$K vanishes are shown as functions of graphene chemical
potential by the four lines from bottom to top for the values of graphene
energy gap $\Delta=0$, 0.1, 0.15, and 0.2~eV, respectively.
}
\end{figure}

In the end of this section, we compute the value of $a_0$, where the Casimir pressure
in dried peptide film vanishes, as a function of the chemical potential of a graphene
sheet for different values of its energy gap. The computational results are shown
in Fig.~\ref{fg3} by the four lines from bottom to top for $\Delta=0$, 0.1, 0.15, and
0.2~eV, respectively. As is seen  in Fig.~\ref{fg3}, with increasing $\mu$ the
quantity $a_0$ decreases. However, an increase of $\Delta$, especially at small $\mu$,
results in significant increase of $a_0$. These results are in agreement with
Fig.~\ref{fg2}(a,b) and show that although graphene coating increases the stability
of peptide film deposited on a metallic substrate, nonzero values of the chemical
potential and the energy gap of graphene influence on this stability in the opposite
directions by increasing and decreasing it, respectively.

\section{The case of doped and gapped graphene coatings}

In this section, we present the results of numerical computations of the Casimir
pressure in a graphene-coated peptide film deposited on an Au substrate in the most
realistic case when peptide contains some fraction of water whereas graphene is
characterized by nonzero values of both the energy gap and chemical potential.

Computations are again performed by Eq.~(\ref{eq2}) where the reflection coefficients
are presented in Eqs.~(\ref{eq4}), (\ref{eq18}) and (\ref{eq19}).
First we consider the peptide film containing $\Phi=0.1$ fraction of water.
The computational results for the Casimir pressure (\ref{eq2}) at $T=300~$K are
presented by the pairs of lines labeled 1 and 2 as functions of film thickness in
the region (a) from 150 to 250~nm and (b) from 250 to 600~nm. The lines in the pair
labeled 1 are computed for graphene sheets with $\mu=0.02~$eV and the  lines in the pair
labeled 2 --- for graphene sheet with $\mu=0.25~$eV. In each pair, the bottom line is for
graphene with $\Delta=0.1~$eV and the top  line is for graphene with $\Delta=0.2~$eV.
Note that graphene coating with relatively low chemical potential $\mu=0.02~$eV was
used in the first experiment on measuring the Casimir interaction in graphene systems
\cite{63}. The results of this experiment were found in good agreement with theoretical
predictions using the polarization tensor of graphene \cite{49}.

\begin{figure}[!b]
\vspace*{-2cm}
\centerline{\hspace*{1cm}
\includegraphics[width=6.50in]{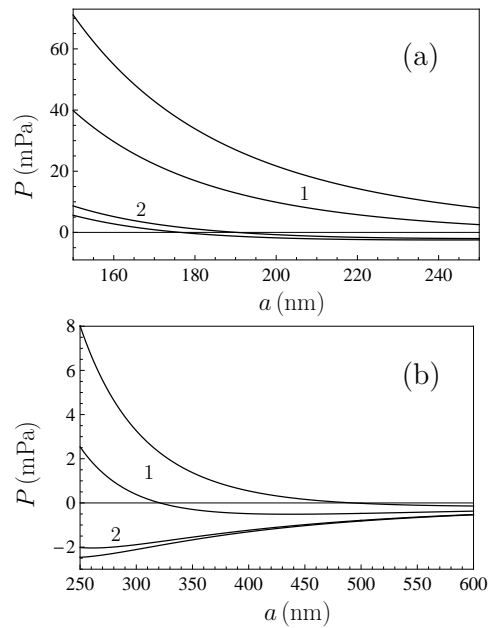}}
\vspace*{-13.cm}
\caption{\label{fg4} The Casimir pressures in peptide film containing 10\% of water
deposited on an Au substrate and coated with a graphene sheet are shown as functions
of film thickness  by the pairs of lines labeled 1 and 2 for the chemical potential
of graphene equal to 0.02 and 0.25~eV, respectively, at $T=300~$K in
the regions of film thickness (a) from 150 to 250~nm and (b) from 250 to 600~nm.
In each pair, the bottom and top lines are for graphene with the energy gap equal
to 0.1 and 0.2~eV, respectively.
}
\end{figure}
Figure~\ref{fg4} allows to trace the joint impact of nonzero energy gap and chemical
potential on the Casimir pressure in peptide film containing 10\% of water.
The pair of lines labeled 1 is always above the pair labeled 2 because it is computed
for smaller chemical potential. Taking into account that in each pair the top line is computed
with a larger value of the energy gap, it is again seen that an increase of the chemical
potential and the energy gap act on the Casimir pressure in opposite directions.
If to compare the pairs of lines labeled 1 and 2, it is seen also that with increasing
$\mu$ an impact of the energy gap on the Casimir pressure is considerably reducing.

As is seen in Fig.~\ref{fg4}(a), the lines of the pair labeled 2 (graphene coating with
$\mu=0.25~$eV) intersect the $a$ axis for the film thicknesses $a_0=176.4~$nm
(graphene coating with $\Delta=0.1~$eV) and  $a_0=190.2~$nm
(graphene coating with $\Delta=0.2~$eV). For thicker peptide films coated with graphene
sheets the Casimir pressure becomes negative which corresponds to attraction.
For graphene coating with $\mu=0.02~$eV this happens for larger film thicknesses.
From Fig.~\ref{fg4}(b) it is seen that the bottom and top lines of the pair labeled 1
(graphene coatings with $\Delta=0.1$ and 0.2~eV, respectively) intersect
the $a$ axis for the film thicknesses $a_0=321$ and 493~nm, respectively.
Only for thicker films the Casimir pressure becomes attractive in this case.

These results can be compared with those presented in Fig.~\ref{fg3} for a dried peptide film.
From Fig.~\ref{fg3} it follows that for $\mu=0.25~$eV the values of $a_0<100~$nm for
any $0\leqslant\Delta\leqslant0.2~$eV. For $\mu=0.02~$eV from  Fig.~\ref{fg3} it is seen
that $a_0=249.4~$nm for a graphene sheet with $\Delta=0.1~$eV and 419.0~nm
for a graphene sheet with $\Delta=0.2~$eV. Thus, in all cases an addition of 10\% of water to
a peptide film substantially increases the minimum film thickness such that the Casimir
pressure in it becomes attractive due to graphene coating.

For comparison purposes, we present the values of the Casimir pressure $P$ (mPa) in peptide
film of 150~nm thickness coated by graphene sheets with different values of $\mu$ and
$\Delta$. The obtained pressures are the following:
39.84 (for $\mu=0.02~$eV, $\Delta=0.1~$eV),
71.10 (for $\mu=0.02~$eV, $\Delta=0.2~$eV),
5.608 (for $\mu=0.25~$eV, $\Delta=0.1~$eV), and
8.657 (for $\mu=0.25~$eV, $\Delta=0.2~$eV).
It is seen that an increase of $\Delta$ with constant $\mu$ increases the value of the
Casimir pressure whereas an increase of $\mu$ with a constant $\Delta$ decreases it in
agreement with the previously obtained results.

\begin{figure}[!t]
\vspace*{-2cm}
\centerline{\hspace*{1cm}
\includegraphics[width=6.50in]{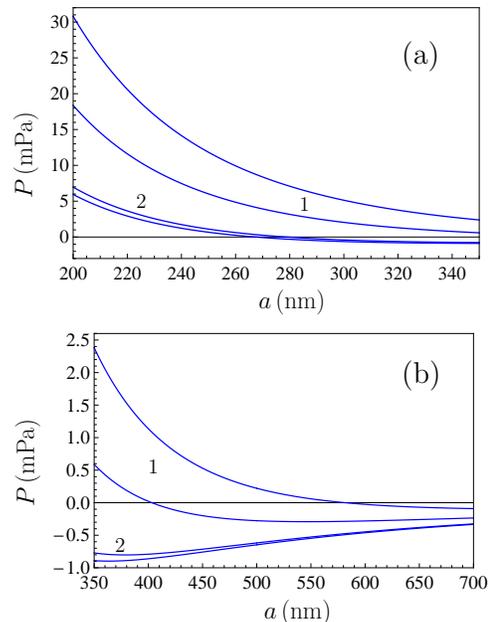}}
\vspace*{-13.1cm}
\caption{\label{fg5} The Casimir pressures in peptide film containing 25\% of water
deposited on an Au substrate and coated with a graphene sheet are shown as functions
of film thickness  by the pairs of lines labeled 1 and 2 for the chemical potential
of graphene equal to 0.02 and 0.25~eV, respectively, at $T=300~$K in
the regions of film thickness (a) from 200 to 350~nm and (b) from 350 to 700~nm.
In each pair, the bottom and top lines are for graphene with the energy gap equal
to 0.1 and 0.2~eV, respectively.
}
\end{figure}

Next we consider the Casimir pressure in peptide film containing $\Phi=0.25$ fraction
of water. Numerical computations were performed in the same way and using the same
parameters of a graphene sheet as discussed above in the case $\Phi=0.1$.
The computational results are presented in Fig.~\ref{fg5}(a,b) similarly to Fig.~\ref{fg4}.
As is seen in Fig.~\ref{fg5}(a), for graphene sheets with $\mu=0.25~$eV, $\Delta=0.1~$eV
and $\mu=0.25~$eV, $\Delta=0.2~$eV (bottom and top lines in the pair labeled 2)
the Casimir pressure vanishes for the films of thickness $a_0=267.5$ and 278.7~nm,
respectively.

{}From Fig.~\ref{fg5}(b) we obtain that for graphene sheets with lower doping
concentration ($\mu=0.02~$eV) the Casimir pressure vanishes
for the peptide film thicknesses $a_0=403.2$ and 581.5~nm for the energy gap of
graphene sheet $\Delta=0.1$ and 0.2~eV, respectively (bottom and top lines in the pair
labeled 1). A comparison with the previously obtained results for peptide films
containing 10\% of water shows that an increase of the percentage of water further
increases the values of film thickness delivering zero value of the Casimir
pressure by means of a graphene coating.

As expected, an increase of the fraction of water in peptide film of the same
thickness results in larger Casimir pressures. Thus, for a film of 150~nm
thickness containing 25\% of water one obtains $P=66.39$, 98.67, 30.00, and 33.02~mPa
for graphene coatings with
$\mu=0.02~$eV, $\Delta=0.1~$eV; $\mu=0.02~$eV, $\Delta=0.2~$eV;
$\mu=0.25~$eV, $\Delta=0.1~$eV; and $\mu=0.25~$eV, $\Delta=0.2~$eV, respectively
(this value of film thickness is not reflected in Fig.~\ref{fg5} aiming to illustrate
the effect of change of sign of the Casimir pressure due to graphene coating).

\begin{figure}[!t]
\vspace*{-2cm}
\centerline{\hspace*{1cm}
\includegraphics[width=6.50in]{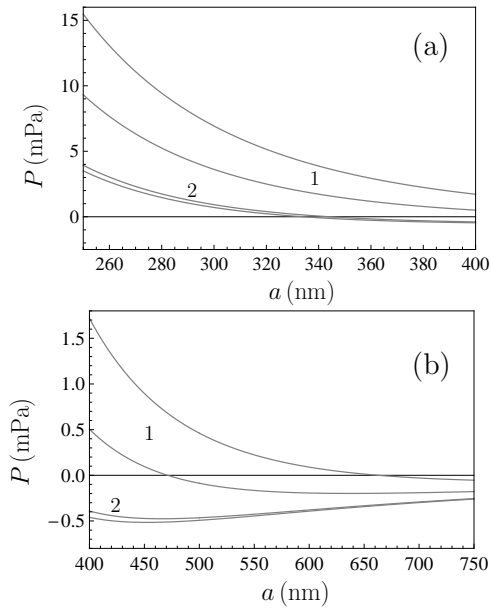}}
\vspace*{-13.1cm}
\caption{\label{fg6} The Casimir pressures in peptide film containing 40\% of water
deposited on an Au substrate and coated with a graphene sheet are shown as functions
of film thickness  by the pairs of lines labeled 1 and 2 for the chemical potential
of graphene equal to 0.02 and 0.25~eV, respectively, at $T=300~$K in
the regions of film thickness (a) from 250 to 400~nm and (b) from 400 to 750~nm.
In each pair, the bottom and top lines are for graphene with the energy gap equal
to 0.1 and 0.2~eV, respectively.
}
\end{figure}

As the last example, we consider the peptide film containing 40\% of water ($\Phi=0.4$).
The computational results for the Casimir pressure are presented in Fig.~\ref{fg6}
in the same way as in Figs.~\ref{fg4} and \ref{fg5}. These results confirm that with
increasing fraction of water in the film the change of sign in the Casimir pressure takes
place for thicker films with the same parameters of graphene coating.
Specifically, from Fig.~\ref{fg6}(a) one obtains that $a_0=334.0~$nm for a graphene
sheet with $\mu=0.25~$eV, $\Delta=0.1~$eV and 344.1~nm for $\mu=0.25~$eV, $\Delta=0.2~$eV
(the pair of lines labeled 2). Similarly, from Fig.~\ref{fg6}(b) one finds $a_0=468.0~$nm for
$\mu=0.02~$eV, $\Delta=0.1~$eV and 662.8~nm for $\mu=0.02~$eV, $\Delta=0.2~$eV
(the pair of lines labeled 1).  For peptide films with $a>a_0$ the Casimir pressure becomes
attractive due to the role of graphene coating and, thus, contributes to the film stability.
\section{Conclusions and discussion}

In the foregoing, we have developed the formalism allowing calculation
of the fluctuation-induced Casimir pressure in the graphene-coated peptide
films deposited on metallic substrates at any temperature. Within this
formalism, metal and peptide are described by the frequency-dependent
dielectric permittivities and graphene --- by the exact polarization
tensor in (2+1)-dimensional space-time found in the framework of
Dirac model. For film thicknesses exceeding 100 nm considered in the
paper, the Matsubara frequencies contributing to the Casimir effect
described by the Lifshitz theory are well in the application region of
the Dirac model. Because of this, the used formalism can be considered
as well justified and based on first principles of thermal quantum
field theory.

The developed formalism was applied to calculate the Casimir pressure
in the graphene-coated dried and containing different fractions of water
peptide films deposited on a metallic (Au) substrate. In so doing the
cases of a pristine graphene, either doped or gapped graphene (which
possesses either nonzero energy gap or nonzero chemical potential),
and also both gapped and doped graphene were considered. The system
of an uncoated peptide film deposited on a metallic substrate deserves
attention because the Casimir pressure in this case turns out to be
positive and, thus, corresponds to a repulsion by making film less
stable. The coating by a graphene sheet is considered as a means to
remedy this defect which may hamper using peptide films of less than
a micrometer thickness deposited on metallic substrates in organic
electronics.

According to our results, the presence of a graphene layer on top of a
peptide film significantly decreases the value of the Casimir pressure in
the film. It is shown that for the films of sufficiently large thickness
$a_0$ (which varies from about 100 nm to several hundred nanometers) the
Casimir pressure vanishes and for thicker films becomes negative by
contributing to their stability. Numerical computations performed for
typical fractions of water in peptide films and representative values
of the energy gap $\Delta$ and chemical potential $\mu$ of graphene
coatings allowed to conclude that with increasing $\mu$ the value
of $a_0$ decreases whereas increase of $\Delta$ and fraction of water
in the film leads to larger values of $a_0$.

These results have simple physical explanation. The point is that in the
configuration of an uncoated peptide film on a metallic substrate all
Matsubara terms in the Lishitz formula (\ref{eq2}) are positive and, thus,
contribute to a repulsion. In the presence of a graphene sheet on top of
peptide film, the reflection coefficient $R_{\rm TM}$ defined in
Eq.~(\ref{eq11}) contains the positive term due to the reflection
coefficient $r_{\rm TM}^{({\rm gr})}$ on a plane between vacuum and graphene and
the negative term due to the reflection coefficient $r_{\rm TM}^{(1,0)}$
on a plane between peptide and vacuum (a contribution of the TE
polarization is much smaller than of the TM one). For $l=0$ the positive
term in $R_{\rm TM}$ defined in Eq.~(\ref{eq18}) is dominant, so that
starting from some film thickness, when the zero-frequency Matsubara
term gives the major contribution to Eq.~(\ref{eq2}), the Casimir pressure
turns out to be negative, i.e., attractive. An increase of $\mu$ and
$\Delta$ leads to an increase and a decrease of the Matsubara term with
$l=0$, respectively, resulting in respective decrease and increase of
$a_0$. In a similar way, an increase of the fraction of water  a
peptide film leads to larger $\ve_0^{(1)}$ and, thus, to larger
magnitudes of the negative term in $R_{\rm TM}(0,y)$ in Eq.~(\ref{eq18}).
The latter in its turn increases the value of $a_0$.

Computations performed in Secs.~III and IV made it possible to determine the
joint action of graphene energy gap and chemical potential, as well as the
fraction of water in a peptide film, on the Casimir pressure and reliably
predict the combination of these parameters which result in the Casimir
attraction favorable for the film stability. Taking into account that
for a peptide film of 100~nm thickness containing 10\% of water at
$T=300~$K the fluctuation-induced free energy contributes from 5\% to
20\% of the total cohesive energy of a film \cite{43}, one can conclude
that the role of fluctuation phenomena in organic electronics is an
important problem which deserves further investigation.

\section*{ACKNOWLEDGMENTS}

The work was of E.~N.~V.\ was supported by the Russian Science Foundation under
grant No.\ 21-72-20029. G.~L.~K.\ and V.~M.~M.\ were partially
supported by the Peter the Great Saint Petersburg Polytechnic
University in the framework of the Russian state assignment for basic research
(project No.\ FSEG-2020-0024).
V.~M.~M.~was partially funded by the Russian Foundation for Basic
Research, grant No.\ 19-02-00453 A.
V.~M.~M.\ was also partially supported by the Russian Government
Program of Competitive Growth of Kazan Federal University.



\begin{thebibliography}{99}
\bibitem{1}
G.~Meller and T.~Grasser (eds.),
{\it Organic Electronics} (Springer, Heidelberg, 2010).
\bibitem{2}
Jingbi You, Letian Dou, Ken Yoshimura, Takehito Kato, Kenichiro Ohya,
Tom Moriarty, Keith Emery, Chun-Chao Chen, Jing Gao, Gang Li, and Yang Yangb,
A polymer tandem solar cell with 10.6\% power conversion efficiency,
Nat. Commun. {\bf 4}, 1446 (2013).
\bibitem{3}
Lingxian Meng, Yamin Zhang, Xiangjian Wan, Chenxi Li, Xin Zhang,
Yanbo Wang, Xin Ke, Zuo Xiao, Liming Ding, Ruoxi Xia, Hin-Lap Yip,
Yong Cao, and Yongsheng Chen,
Organic and solution-processed tandem solar cells with 17.3\% efficiency,
Science {\bf 361}, 1094 (2018).
\bibitem{4}
C.~D.~Dimitrakopoulos and P.~R.~L.~Malenfant,
Organic thin film transistors for large area electronics,
Adv. Mater. {\bf 14}, 99 (2002).
\bibitem{4a}
M.~E.~Gershenson, V.~Podzorov, and A.~F.~Morpurgo,
{\it Colloquium}: Electronic transport in single-crystal organic transistors,
Rev. Mod. Phys. {\bf 78}, 973 (2006).
\bibitem{5}
Chun-Yi Lee, Jenn-Chang Hwang, Yu-Lun Chueh, Ting-Hao Chang, Yi-Yun Cheng,
and Ping-Chiang Lyu,
Hydrated bovine serum albumin as the gate dielectric material for organic
field-effect transistors,
Org. Electr. {\bf 14}, 2645 (2013).
\bibitem{6}
Mingchao Ma, Xinjun Xu, Leilei Shi, and Lidong Li,
Organic field-effect transistors with a low driving voltage using albumin
as the dielectric layer,
RSC Advances {\bf 4}, 58720 (2014).
\bibitem{7}
Hiroki Uoyama, Kenichi Goushi, Katsuyuki Shizu, Hiroko Nomura, and
Chihaya Adachi,
Highly efficient organic light-emitting diodes from delayed fluorescence,
Nature {\bf 492}, 234 (2012).
\bibitem{8}
Qisheng Zhang, Bo Li, Shuping Huang, Hiroko Nomura, Hiroyuki Tanaka, and
Chihaya Adachi,
Efficient blue organic light-emitting diodes employing thermally
activated delayed fluorescence,
Nat. Photon. {\bf 8}, 326 (2014).
\bibitem{9}
M.~Natesan and R.~G.~Ulrich,
Protein microarrays, and biomarkers of infection disease,
Int. J. Mol. Sci. {\bf 11}, 5165 (2010).
\bibitem{10}
C.-K.~Chou, N.~Jing, H.~Yamaguchi, P.-H.~Tsou, H.-H.~Lee, C.-T.~Chen,
Y.-N.~Wang, S.~Hong, C.~Su, J.~Kameoka, and M.-C.~Hung,
Rapid detection of two-protein interaction with a single fluorophore by
using a microfluidic device,
Analyst {\bf 135}, 2907 (2010).
\bibitem{11}
H.~Chandra, P.~J.~Reddy, and S.~Srivastava,
Protein microarrays and novel detection platforms,
Expert Rev. Proteomics {\bf 8}, 61 (2011).
\bibitem{12}
Bin Zheng, D.~T. Haynie, Hua Zhong, K. Sabnis, V. Surpuriya, N. Pargaonkar,
G. Sharma, and K. Vistakula,
Design of peptides for thin films, coatings and microcapsules for
applications in biotechnology,
J. Biomater. Sci., Polymer Edit. {\bf 16}, 285 (2005).
\bibitem{13}
B.~Li, D.~T.~Haynie, N.~Palath, and D.~Janisch,
Nano-Scale Biomimetics: Fabrication and Optimization of Stability of
Peptide-Based Thin Films,
J. Nanosci. Nanotech. {\bf 5}, 2042 (2005).
\bibitem{14}
M.~Righi, G.~L.~Puleo, I.~Tonazzini, G.~Giudetti, M.~Cecchini, and S.~Micera,
Peptide-based coatings for flexible implantable neural interfaces,
Sci. Reports {\bf 8}, 502 (2018).
\bibitem{15}
T.~Guterman and E.~Gazit,
Toward peptide-based bioelectronics: reductionist design of conductive
pili mimetics,
Bioelectron. Med. (Lond.) {\bf 1}, 131 (2018).
\bibitem{16}
J.~Yu, J.~R.~Horsley, and A.~D.~Abell,
Peptides as Bio-Inspired Electronic Materials: An Electrochemical and
First-Principles Perspective,
Acc. Chem. Res. {\bf 51}, 2237 (2018).
\bibitem{17}
S.~S.~Panda, H.~E.~Katz, and J.~D.~Tovar,
Solid-state electrical applications of protein and peptide based nanomaterials,
Chem. Soc. Rev. {\bf 47}, 3640 (2018).
\bibitem{18}
A.~Arul, S.~Sivagnanam, A.~Dey, O.~Mukherjee, S.~Ghosh, and P.~Das,
The design and development of short peptide-based novel smart materials to
prevent fouling by the formation of non-toxic and biocompatible coatings,
RSC Advances {\bf 10}, 13420 (2020).
\bibitem{19}
U.~Haas, A.~Haase, V.~Satzinger, H.~Pichler, G.~Leising, G.~Jakopic,
B.~Stadlober, R.~Houbertz, G.~Domann, and A.~Schmitt,
Hybrid polymers as tunable and directly-patternable gate dielectrics in
organic thin-film transistors,
Phys. Rev. B {\bf 73}, 235339 (2006).
\bibitem{20}
L.~Romaner, G.~Heimel, J.-L.~Br\'{e}das, A.~Gerlach, F.~Schreiber,
R.~L.~Johnson, J.~Zegenhagen, S.~Duhm, N.~Koch, and E.~Zojer,
Impact of Bidirectional Charge Transfer and Molecular Distortions on
the Electronic Structure of a Metal-Organic Interface,
Phys. Rev. Lett. {\bf 99}, 256801 (2007).
\bibitem{21}
H.~Yamane, A.~Gerlach, S.~Duhm, Y.~Tanaka, T.~Hosokai, Y.~Y.~Mi, J.~Zegenhagen, 
N.\ Koch, H.\ Seki, and F.\ Schreiber,
Site-Specific Geometric and Electronic Relaxations at 
Organic-Metal Surfaces,
Phys. Rev. Lett. {\bf 105}, 046103 (2010).
\bibitem{22}
M.~G.~Helander, M.~T.~Greiner, Z.~B.~Wang, and Z.~H.~Lu,
 Effect of electrostatic screening on apparent shifts in photoemission
spectra near metal/organic interfaces,
Phys. Rev. B {\bf 81}, 153308 (2010).
\bibitem{23}
A.~V.~Nenashev, S.~D.~Baranovskii, M.~Wiemer, F.~Jansson, R.~\"{O}sterbacka,
A.~V.~Dvurechenskii, and F.~Gebhard,
Theory of exciton dissociation at the interface between a conjugated polymer
and an electron acceptor,
Phys. Rev. B {\bf 84}, 035210 (2011).
\bibitem{24}
S. Ciuchi and S. Fratini,
Electronic transport and quantum localization effects in organic
semiconductors,
Phys. Rev. B {\bf 86}, 245201 (2012).
\bibitem{25}
S.~Pittner, D.~Lehmann, D.~R.~T.~Zahn, and V.~Wagner,
Charge transport analysis of poly(3-hexylthiophene) by electroreflectance
spectroscopy,
Phys. Rev. B {\bf 87}, 115211 (2013).
\bibitem{26}
Shota Ono and Kaoru Ohno,
Minimal model for charge transfer excitons at the dielectric interface.
Phys. Rev. B {\bf 93}, 121301(R) (2016).
\bibitem{27}
Yuhan Zhang, Jingsi Qiao, Si Gao, Fengrui Hu, Daowei He, Bing Wu, Ziyi Yang,
Bingchen Xu, Yun Li, Yi Shi, Wei Ji, Peng Wang, Xiaoyong Wang, Min Xiao,
Hangxun Xu, Jian-Bin Xu, and Xinran Wang,
Probing Carrier Transport and Structure-Property Relationship of Highly
Ordered Organic Semiconductors at the Two-Dimensional Limit,
Phys. Rev. Lett. {\bf 116}, 016602 (2016).
\bibitem{28}
K.~Stallberg, A.~Namgalies, and U.~H\"{o}fer,
Photoluminescence study of the exciton dynamics at PTCDA/noble-metal
interfaces,
Phys. Rev. B {\bf 99}, 125410 (2019).
\bibitem{29}
C.~Metzger, M.~Graus, M.~Grimm, G.~Zamborlini, V.~Feyer, M.~Schwendt,
D.~L\"{u}ftner, P.~Puschnig, A.~Sch\"{o}ll, and F.~Reinert,
Plane-wave final state for photoemission from nonplanar molecules at a
metal-organic interface,
Phys. Rev. B {\bf 101}, 165421 (2020).
\bibitem{30}
V.~A.~Parsegian,
{\it Van der Waals Forces: A Handbook for Biologists, Chemists, Engineers,
and Physicists}
(Cambridge University Press, Cambridge, 2005).
\bibitem{31}
M.~Bordag, G.~L.~Klimchitskaya, U.~Mohideen, and V.\ M.\ Mostepanenko,
{\it Advances in the Casimir Effect}
(Oxford University Press, Oxford, 2015).
\bibitem{32}
E.~M.~Lifshitz and L.~P.~Pitaevskii,
{\it Statistical Physics, Pt. II}
(Pergamon Press, Oxford, 1980).
\bibitem{33}
V.~A.~Parsegian and B.~W.~Ninham,
Application of the Lifshitz theory to the calculation of
van der Waals forces across thin lipid films,
Nature {\bf 224}, 1197 (1972).
\bibitem{34}
S.~Nir,
Van der Waals interactions between surfaces of biological interest,
Progr. Surf. Sci. {\bf 8}, 1 (1976).
\bibitem{35}
C.~M.~Roth, B.~L.~Neal, and A.~M.~Lenhoff,
Van der Waals interactions involving proteins,
Biophys. J. {\bf 70}, 977 (1996).
\bibitem{36}
Bing-Sui Lu and R. Podgornik,
Effective interactions between fluid membranes,
Phys. Rev. E {\bf 92}, 022112 (2015).
\bibitem{37}
G.~L.~Klimchitskaya and V.~M. ~Mostepanenko,
Casimir free energy of metallic films: Discriminating between Drude and plasma
model approaches,
Phys. Rev. A {\bf 92}, 042109 (2015).
\bibitem{38}
G.~L.~Klimchitskaya and V.~M.~Mostepanenko,
Casimir and van der Waals energy of anisotropic atomically thin metallic films,
Phys. Rev. B {\bf 92}, 205410 (2015).
\bibitem{39}
G.~L.~Klimchitskaya and V.~M.~Mostepanenko,
Casimir free energy and pressure for magnetic metal films,
Phys. Rev. B {\bf 94}, 045404 (2016).
\bibitem{40}
G.~L.~Klimchitskaya and V.~M.~Mostepanenko,
Characteristic properties of the Casimir free energy for metal films deposited
on metallic plates,
Phys. Rev. A {\bf 93}, 042508 (2016).
\bibitem{41}
G.~L.~Klimchitskaya and V.~M.~Mostepanenko,
Low-temperature behavior of the Casimir free energy and entropy of metallic
films,
Phys. Rev. A {\bf 95}, 012130 (2017).
\bibitem{42}
G.~L.~Klimchitskaya and V.~M.~Mostepanenko,
Casimir free energy of dielectric films: Classical limit, low-temperature
behavior and control,
J. Phys.: Condens. Matter {\bf 29}, 275701 (2017).
\bibitem{43}
M.~A.~Baranov, G.~L.~Klimchitskaya, V.~M.~Mostepanenko, and E.~N.~Velichko,
Fluctuation-induced free energy of thin peptide films,
Phys. Rev. E {\bf 99}, 022410 (2019).
\bibitem{44}
G.~L\"{o}ffler, H.~Schreiber, and O.~Steinhauser,
Calculation of the Dielectric Properties of a Protein and its Solvent:
Theory and a Case Study,
J. Mol. Biol. {\bf 270}, 520 (1997).
\bibitem{45}
P.~Adhikari, A.~M.~Wen, R.~H.~French, V.~A.~Parsegian, N.~F.~Steinmetz,
R.~Podgornik, and W.-Y.~Ching,
Electronic Structure, Dielectric Response, and Surface Charge Distribution of
RGD (1FUV) Peptide,
Sci. Reports. {\bf 4}, 5605 (2014).
\bibitem{46}
E.~N.~Velichko, M.~A.~Baranov, and V.~M.~Mostepanenko,
Change of sign in the Casimir interaction of peptide films deposited
on a dielectric substrate,
Mod. Phys. Lett. A {\bf 35}, 2040020 (2020).
\bibitem{47}
V.~M.~Mostepanenko, E.~N.~Velichko, and M.~A.~Baranov,
Role of Electromagnetic Fluctuations in Organic Electronics,
J. Electr. Sci. Tech. {\bf 18}, 100023 (2020).
\bibitem{48}
G.~L.~Klimchitskaya, V.~M.~Mostepanenko, and E.~N.~Velichko,
Effect of increased stability of peptide-based coatings in the Casimir
regime via nanoparticle doping,
Phys. Rev. B {\bf 102}, 161405(R) (2020).
\bibitem{49}
G.~L.~Klimchitskaya, U.~Mohideen, and V.~M.~Mostepanenko,
Theory of the Casimir interaction for graphene-coated substrates
 using the polarization tensor and comparison with experiment,
Phys. Rev. B {\bf 89}, 115419 (2014).
\bibitem{50}
A.~H.~Castro Neto, F.~Guinea, N.~M.~R.~Peres, K.~S.~Novoselov,
and A.~K.~Geim,
The electronic properties of graphene,
Rev. Mod. Phys. {\bf 81}, 109 (2009).
\bibitem{51}
M.~Bordag, I.~V.~Fialkovsky, D.~M.~Gitman, and
D.~V.~Vassilevich,
Casimir interaction between a perfect conductor and graphene
described by the Dirac model,
{Phys. Rev. B} {\bf 80}, 245406 (2009).
\bibitem{52}
I.~V.~Fialkovsky, V.~N.~Marachevsky, and
D.~V.~Vassilevich,
Finite-temperature Casimir effect for graphene,
{Phys. Rev. B} {\bf 84}, 035446 (2011).
\bibitem{53}
G.~Bimonte, G.~L.~Klimchitskaya, and V.~M.~Mostepanenko,
Thermal effect in the Casimir force for graphene and graphene-coated
substrates: Impact of nonzero mass gap and chemical potential,
Phys. Rev. B {\bf 96}, 115430 (2017).
\bibitem{54}
C.~Henkel, G.~L.~Klimchitskaya, and V.~M.~Mostepanenko,
Influence of chemical potential on the Casimir-Polder interaction
between an atom and gapped graphene or graphene-coated substrate,
Phys. Rev. A {\bf 97}, 032504 (2018).
\bibitem{55}
M.~Bordag, G.~L.~Klimchitskaya, V.~M.~Mostepanenko, and V.~M.~Petrov,
Quantum field theoretical description for the reflectivity of graphene,
Phys. Rev. D {\bf 91}, 045037 (2015); {\bf 93}, 089907(E) (2016).
\bibitem{56}
M.~Bordag, I.~Fialkovskiy, and D.~Vassilevich,
Enhanced Casimir effect for doped graphene,
Phys. Rev. B {\bf 93}, 075414 (2016);
{\bf 95}, 119905(E) (2017).
\bibitem{57}
G.~L.~Klimchitskaya and V.~M.~Mostepanenko,
Origin of large thermal effect in the Casimir interaction between two
graphene sheets,
Phys. Rev. B {\bf 91}, 174501 (2015).
\bibitem{58}
{\it Handbook of Optical Constants of Solids},
ed. E.~D.~Palik (Academic, New York, 1985).
\bibitem{59}
G.~L.~Klimchitskaya, U.~Mohideen, and V.~M.~Mostepanenko,
The Casimir force between real materials: Experiment and theory,
Rev. Mod. Phys. {\bf 81}, 1827 (2009).
\bibitem{60}
L.~M.~Woods, D.~A.~R.~Dalvit, A.~Tkatchenko, P.~Rodriguez-Lopez,
A.~W.~Rodriguez, and R.~Podgornik,
Materials perspective on Casimir and van der Waals interactions,
Rev. Mod. Phys. {\bf 88}, 045003 (2016).
\bibitem{60a}
G.~Bimonte, B.~Spreng, P.~A.~Maia Neto, G.-L.~Ingold, G.~L.~Klimchitskaya,
V.~M.~Mostepanenko, and R.~S.~Decca,
Measurement of the Casimir Force between 0.2 and 8 �m: Experimental
Procedures and Comparison with Theory,
Universe {\bf 7}, 93 (2021).
\bibitem{61}
V.~M.~Mostepanenko,
Casimir Puzzle and Casimir Conundrum: Discovery and Search for Resolution,
Universe {\bf 7}, 84 (2021).
\bibitem{62}
P.~Shemella and S.~K.~Nayak,
Electronic structure and band-gap modulation of graphene via substrate
surface chemistry,
Appl. Phys. Lett. {\bf 94}, 032101 (2009).
\bibitem{63}
A.~A.~Banishev, H.~Wen, J.~Xu, R.~K.~Kawakami, G.~L.~Klimchitskaya,
V.~M.~Mostepanenko, and U.~Mohideen,
Measuring the Casimir force gradient from graphene on a SiO$_2$ substrate
Phys. Rev. B {\bf 87}, 205433 (2013).

\end{thebibliography}
\end{document}